\documentclass[aps,prc,showpacs,twocolumn,superscriptaddress,10pt]{revtex4}
\usepackage{amsmath} 
\usepackage{amssymb}
\usepackage{graphicx}
\usepackage{graphics}
\usepackage{dcolumn}
\usepackage{bm}
\usepackage{epsfig}
\usepackage{color}
\begin{document}
\title{Measurements of identified particles at intermediate transverse momentum in the STAR experiment from Au+Au collisions at $\sqrt{s_{NN}}=200~\mathrm{GeV}$}
\affiliation{Argonne National Laboratory, Argonne, Illinois 60439}
\affiliation{University of Birmingham, Birmingham, United Kingdom}
\affiliation{Brookhaven National Laboratory, Upton, New York 11973}
\affiliation{California Institute of Technology, Pasadena, California 91125}
\affiliation{University of California, Berkeley, California 94720}
\affiliation{University of California, Davis, California 95616}
\affiliation{University of California, Los Angeles, California 90095}
\affiliation{Carnegie Mellon University, Pittsburgh, Pennsylvania 15213}
\affiliation{Creighton University, Omaha, Nebraska 68178}
\affiliation{Nuclear Physics Institute AS CR, 250 68 \v{R}e\v{z}/Prague, Czech Republic}
\affiliation{Laboratory for High Energy (JINR), Dubna, Russia}
\affiliation{Particle Physics Laboratory (JINR), Dubna, Russia}
\affiliation{University of Frankfurt, Frankfurt, Germany}
\affiliation{Institute of Physics, Bhubaneswar 751005, India}
\affiliation{Indian Institute of Technology, Mumbai, India}
\affiliation{Indiana University, Bloomington, Indiana 47408}
\affiliation{Institut de Recherches Subatomiques, Strasbourg, France}
\affiliation{University of Jammu, Jammu 180001, India}
\affiliation{Kent State University, Kent, Ohio 44242}
\affiliation{Institute of Modern Physics, Lanzhou, P.R. China}
\affiliation{Lawrence Berkeley National Laboratory, Berkeley, California 94720}
\affiliation{Massachusetts Institute of Technology, Cambridge, MA 02139-4307}
\affiliation{Max-Planck-Institut f\"ur Physik, Munich, Germany}
\affiliation{Michigan State University, East Lansing, Michigan 48824}
\affiliation{Moscow Engineering Physics Institute, Moscow Russia}
\affiliation{City College of New York, New York City, New York 10031}
\affiliation{NIKHEF and Utrecht University, Amsterdam, The Netherlands}
\affiliation{Ohio State University, Columbus, Ohio 43210}
\affiliation{Panjab University, Chandigarh 160014, India}
\affiliation{Pennsylvania State University, University Park, Pennsylvania 16802}
\affiliation{Institute of High Energy Physics, Protvino, Russia}
\affiliation{Purdue University, West Lafayette, Indiana 47907}
\affiliation{Pusan National University, Pusan, Republic of Korea}
\affiliation{University of Rajasthan, Jaipur 302004, India}
\affiliation{Rice University, Houston, Texas 77251}
\affiliation{Universidade de Sao Paulo, Sao Paulo, Brazil}
\affiliation{University of Science \& Technology of China, Hefei 230026, China}
\affiliation{Shanghai Institute of Applied Physics, Shanghai 201800, China}
\affiliation{SUBATECH, Nantes, France}
\affiliation{Texas A\&M University, College Station, Texas 77843}
\affiliation{University of Texas, Austin, Texas 78712}
\affiliation{Tsinghua University, Beijing 100084, China}
\affiliation{Valparaiso University, Valparaiso, Indiana 46383}
\affiliation{Variable Energy Cyclotron Centre, Kolkata 700064, India}
\affiliation{Warsaw University of Technology, Warsaw, Poland}
\affiliation{University of Washington, Seattle, Washington 98195}
\affiliation{Wayne State University, Detroit, Michigan 48201}
\affiliation{Institute of Particle Physics, CCNU (HZNU), Wuhan 430079, China}
\affiliation{Yale University, New Haven, Connecticut 06520}
\affiliation{University of Zagreb, Zagreb, HR-10002, Croatia}

\author{J.~Adams}\affiliation{University of Birmingham, Birmingham, United Kingdom}
\author{M.M.~Aggarwal}\affiliation{Panjab University, Chandigarh 160014, India}
\author{Z.~Ahammed}\affiliation{Variable Energy Cyclotron Centre, Kolkata 700064, India}
\author{J.~Amonett}\affiliation{Kent State University, Kent, Ohio 44242}
\author{B.D.~Anderson}\affiliation{Kent State University, Kent, Ohio 44242}
\author{M.~Anderson}\affiliation{University of California, Davis, California 95616}
\author{D.~Arkhipkin}\affiliation{Particle Physics Laboratory (JINR), Dubna, Russia}
\author{G.S.~Averichev}\affiliation{Laboratory for High Energy (JINR), Dubna, Russia}
\author{S.K.~Badyal}\affiliation{University of Jammu, Jammu 180001, India}
\author{Y.~Bai}\affiliation{NIKHEF and Utrecht University, Amsterdam, The Netherlands}
% ...

\collaboration{STAR Collaboration}\noaffiliation


\homepage{www.star.bnl.gov}\noaffiliation






\author{A.~Braem}\affiliation{CERN, Switzerland}
\author{M.~Davenport}\affiliation{CERN, Switzerland}
\author{G.~De~Cataldo}\affiliation{INFN, Sez. Di Bari, Bari, Italy}
\author{D.~Di~Bari}\affiliation{INFN, Sez. Di Bari, Bari, Italy}
\author{A.~Di~Mauro}\affiliation{CERN, Switzerland}
\author{G.J.~Kunde}\affiliation{Los Alamos National Laboratory, Los Alamos, New Mexico 87545}
\author{P.~Martinengo}\affiliation{CERN, Switzerland}
\author{E.~Nappi}\affiliation{INFN, Sez. Di Bari, Bari, Italy}
\author{G.~Paic}\affiliation{Instituto de Ciencias Nucleares, UNAM, Mexico}
\author{E.~Posa}\affiliation{INFN, Sez. Di Bari, Bari, Italy}
\author{F.~Piuz}\affiliation{CERN, Switzerland}
\author{E.~Schyns}\affiliation{CERN, Switzerland}

\collaboration{STAR-RICH Collaboration}\noaffiliation
\date{\today}
\begin{abstract}
Data for Au+Au collisions at $\sqrt{s_{NN}}=200~\mathrm{GeV}$ are analyzed to determine the ratios of identified hadrons ($\pi$, $K$, $p$, $\Lambda$) as  functions of collision centrality and transverse momentum ($p_T$).  We find that ratios of anti-baryon to baryon yields are independent of $p_T$ up to $5~\mathrm{GeV}/c$, a result inconsistent with results of theoretical pQCD calculations that predict a decrease due to a stronger contribution from valence quark scattering.  For both strange and non-strange species, strong baryon enhancements relative to meson yields are observed as a function of collision centrality in the intermediate $p_T$ region, leading to $p/\pi$ and $\Lambda$/K ratios greater than unity.  The increased $p_T$ range offered by the $\Lambda$/K$^{0}_{S}$ ratio allows a test of the  applicability of various models developed for the  intermediate $p_{T}$ region.
The physics implications of these measurements are discussed with regard to different theoretical models.
\end{abstract}
\maketitle
\section{Introduction}
\label{lab:Introduction}
The Relativistic Heavy Ion Collider (RHIC) at Brookhaven National Laboratory (BNL) is dedicated to the production and study of strongly interacting matter under extreme conditions of temperature and density, formed when nuclei at very high energies collide.
Lattice QCD predicts for such extreme conditions that the strongly interacting matter will undergo a phase transition leading to a system where the degrees of freedom are partonic~\cite{Ref:LQCD1,Ref:LQCD2,Ref:LQCD3}.
Experimentally, however, it is not possible to observe such a partonic system, but rather one observes a system that has evolved from the partonic system and whose degrees of freedom are hadronic. These hadronic data must then be used to 
infer information about the original partonic system and its evolution.

It has been shown that a hydrodynamic description of relativistic heavy-ion collisions reproduces the main characteristics of soft particle production in the region of relatively low transverse momentum ($\sim p_T < 2~\mathrm{GeV}/c$)~\cite{Ref:HydroModels}.  At high momentum transfer
($\sim p_T > 5~\mathrm{GeV}/c$),
parton-parton scattering and subsequent string fragmentation should dominate particle production, and one expects theoretical pQCD calculations to describe the data arising from hard processes.  The intermediate transverse momentum region ($\sim 2 < p_T < 5~\mathrm{GeV}/c$) sits tacitly between these two extremes, and 
here one might expect data to reflect 
an interplay between both mechanisms rather than a simple superposition; not only could parton fragmentation be influenced by the surrounding matter at high $p_T$~\cite{Ref:MedFf}, but non-perturbative effects might be observed just above the region where the hydrodynamical description can successfully reproduce the measurements~\cite{Ref:HydroSTARData}.
Indeed, it has been suggested that at top RHIC energies, specific non-perturbative mechanisms might exist at intermediate $p_T$~\cite{Ref:Fries_PRC,Ref:PrcSoftQ03,Ref:PrcCoalGKL,Ref:PrcCoalHwa}.
In this paper we present new data which extend to the upper bound of the intermediate $p_{T}$ region and make comparisons of the data with model predictions which incorporate a variety of mechanisms.

The paper is outlined as follows. Section II provides a brief description of the experiment which took data for Au+Au collisions at $\sqrt{s_{NN}}= 200~\mathrm{GeV}$. The details of the analysis, using data from a ring imaging Cherenkov detector and a weak decay topological method, are presented in Section III.  The analysis allows charged particle identification ($p$, $\bar{p}$, K$^+$, K$^-$, $\pi^+$ and $\pi^-$) and neutral hadron reconstruction ($\Lambda$ and $K^{0}_{S}$) in the intermediate $p_T$ region.  The results ($p_T$ spectra, anti-baryon over baryon ratios and baryon over meson ratios) are presented in Section IV. Comparisons of data with  available model predictions are made in Section V.
Finally, Section VI contains a summary and some conclusions.\\ 
\section{Experiment}
During Run 2 at RHIC (years 2001--02) data were taken using the STAR detector~\cite{Ref:NimStar} for Au+Au collisions at 200 GeV per nucleon pair.
For tracking and momentum determination in the fiducial volume, STAR's magnetic field was $0.5~\mathrm{T}$.
For charged particle identification (PID), both the full azimuthal coverage of the Time Projection Chamber (TPC)~\cite{Ref:NimTpc} and a Ring Imaging Cherenkov detector (RICH)~\cite{Ref:HmpidAlice} were used. The latter has smaller acceptance than the TPC but has very good particle identification capabilities in the intermediate momentum range.\\

Two triggers were used to record data. The first, the minimum bias trigger, required a coincidence among two Zero-Degree-Calorimeters (ZDCs) located $\simeq 18~\mathrm{m}$ upstream and $\simeq 18~\mathrm{m}$ downstream from the center of STAR and very close to the beam axis ($\Theta<2~\mathrm{mrad}$)~\cite{Ref:ZDCs}. The ZDCs detect spectator neutrons --- neutrons in the colliding Au nuclei that do not participate in the Au+Au collisions.  The second trigger, a charged particle multiplicity or centrality trigger, required, in addition to ZDC coincidences, hits in STAR's Central Trigger Barrel (CTB).  The CTB is an array of scintillator slats that surround the TPC at a radius of $2~\mathrm{m}$. The CTB signal is proportional to the charged particle multiplicity in the pseudo-rapidity interval $|\eta|<1$. Use of the CTB in the trigger allowed the selection of events according to their centrality class, up to approximately the top [0-5]$\%$ of the inelastic cross-section~\cite{Ref:NimTriggerSTAR}.
In addition, for the highest luminosities and data taking rates, the average number of high transverse momentum particles within an event was enhanced through fast online reconstruction for particles hitting the RICH detector~\cite{Ref:NimL3}.

\section{Analysis}
Data for  $1.6\times 10^{6}$ Au+Au collisions were recorded using the minimum bias trigger, while the centrality trigger was used to record another $1.6\times 10^{6}$ events.
Tracks were reconstructed in the TPC after an off-line selection on the position of the primary vertex along the beam axis ($|z|<25~\mathrm{cm}$, where $z=0$ is the center of STAR) this  was required to avoid selection bias when using ZDC timing only. This vertex cut also limited acceptance bias due to the CTB response for selecting high multiplicity events with the central trigger. The central and minimum bias samples were divided into centrality classes.
Data for the four classes [0--5\%], [20--40\%], [40--60\%] and [60--80\%] are presented and discussed in this paper. 
The most central class contained $800~\mathrm{k}$ events, while each of the other classes contained $325~\mathrm{k}$ events.   Use of the RICH detector provided a smaller subset of  events, approximately 80$\%$ of each sample.
\subsection{Pion, kaon and proton identification with the RICH detector}
The STAR Ring Imaging Cherenkov detector (RICH) is $1~\mathrm{m}^2$ in area and was positioned at central rapidity  a radial distance of $2~\mathrm{m}$ from the beamline.  It\textcolor{blue}{'}s azimuthal coverage was $\Delta \phi <20^\circ$ which corresponds to approximately 2$\%$ of the total TPC acceptance~\cite{Ref:NimRich}.\\
The momentum of each charged particle was determined from the curvature of its reconstructed helical track in the TPC. After reconstruction in the TPC fiducial volume, some tracks were extrapolated to the RICH. The corresponding minimum ionizing particle ({\sc mip}) first crosses the Cherenkov photon radiating medium.
The {\sc mip} radiates photons while going through this $1~\mathrm{cm}$ thick liquid (minimum for a normal incident angle). A $0.5~\mathrm{cm}$ quartz window with very good transmission for created UV-photons~\cite{Ref:NimRich} separates the radiating medium from a proximity gap of $8~\mathrm{cm}$. Both {\sc mip}s and induced photons reach the photo\textcolor{red}{-}detector consisting of a multi-wire proportional chamber with a CsI layer deposited on the pad cathode surface for photo\textcolor{red}{-}conversion. 
Rings are reconstructed from the photo-electron cluster locations due to the Cherenkov light and  the central cluster due to the {\sc mip}. For each particle, a Cherenkov angle is computed using the ring and track information at the radiator; then the relativistic velocity, $\beta$, is obtained using the Cherenkov angle once the momentum is known.\\
\begin{table}
\begin{ruledtabular}
\begin{tabular}{lcl}
 TPC Geometrical Selections &   & Value       \\
\hline
 $p_{T}$                    &$>$& $1.5~$GeV/$c$ \\
 $|\eta|$                   &$<$& $0.20$      \\
 DCA to primary vertex      &$<$& $0.75~$cm   \\
 Fit hits $^{\rm a}$        &$>$& $20$        \\
 Last hit radial distance $^{\rm b}$  &$>$& $140~$cm    \\
\hline
 Ring Imaging Selections    &   & Value       \\
\hline
 MIP cluster charge         &$>$& $150$ {\sc adc} \\
 Ring photons               &$>$& $1$         \\
 Incident angle             &$<$& $17^{0}$    \\  
 Residual                   &$<$& $0.5~$cm    \\
\hline
$^{\rm a}$ maximum number of hits is $45$; &  & \\
$^{\rm b}$ outer radius of the {\sc tpc} is $200$~cm. &  & \\
\end{tabular}
\caption{Values of the geometrical cut variables used to select the TPC tracks entering the RICH detector and values of the cut parameters for the RICH analysis.}
\label{Tab:rich_cuts}
\end{ruledtabular}
\end{table}

\begin{center}
\begin{figure}[ht!]
 \includegraphics[width=0.48\textwidth]{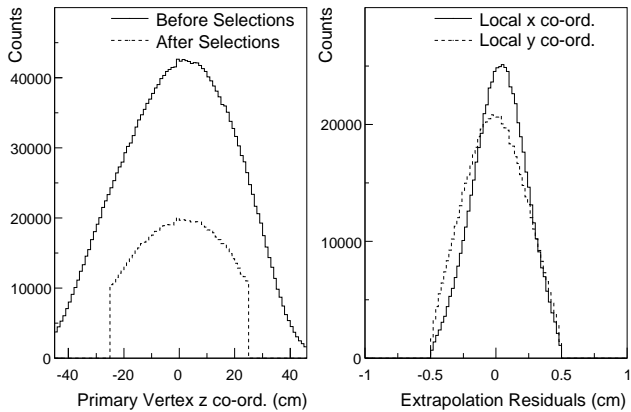} 
\caption{Left panel: The distribution of the primary vertex location along the beam axis before (solid) and after (dashed) centrality and trigger selections. Right panel: Residuals of TPC extrapolated tracks and center of {\sc mip} cluster. Both figures are for the most central event sample. The small offset for the residuals along the x co-ordinate of the RICH plane is due  to a systematic shift in the mean position of the primary vertex.}
\label{Fig:RichRecoRes}
\end{figure}
\end{center}

Primary tracks that are reconstructed in the TPC and within the acceptance of the RICH ($|\eta|<0.2$) are selected with a distance of closest approach (DCA) to the primary vertex $<~0.75~\mathrm{cm}$.  This selection increases the probability that tracks originate from the interaction vertex (primary tracks) rather than from secondary decay products or particles produced via interactions in detector material. Other selections are required in order to have sufficient track quality before extrapolation to the RICH detector: a minimum of 20 hits in the TPC (out of a maximum of 45) with the most external one at a radial distance of at least $140~\mathrm{cm}$.  
\begin{center}
\begin{figure}[ht!]
\includegraphics[width=0.48\textwidth]{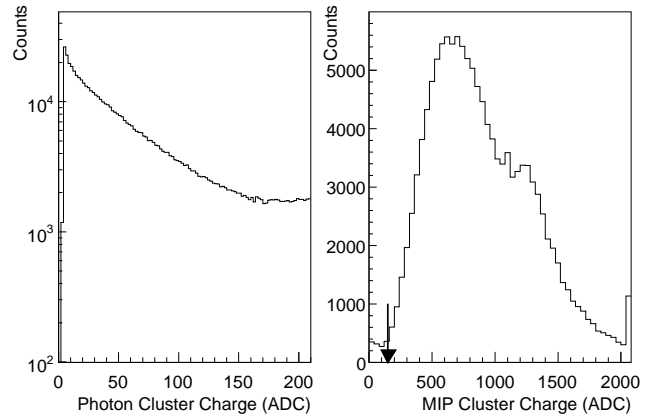} 
\caption{Distributions of cluster charge for all the detected photons (left panel) and for the {\sc mip} after pattern recognition (right panel). The final value for the {\sc mip} selection is indicated by an arrow.}
\label{Fig:RichRecoCharge}
\end{figure}
\end{center}
The energy deposited on the pad plane, larger for a {\sc mip}, can be used to distinguish charged particles from photons.  A cluster charge of more than $150~\mathrm{ADC}$ counts, as shown in Fig.~\ref{Fig:RichRecoCharge}, identifies the {\sc mip} without ambiguity. Its centroid defines the center for the ring recognition algorithm.
It is possible that a ring can be determined by this center and a single photon only (cluster charge $<150~\mathrm{ADCs}$). 
Ring identification is improved by requiring a small incident angle between the detector axis and the direction of the track ($<17^\circ$). 
A comparison of the extrapolated TPC track and the center of the cluster from the {\sc mip}  leads to residual distributions which depend on the position of the interaction vertex along the beam axis.  The related widths are $0.17~\mathrm{cm}$ and $0.20~\mathrm{cm}$ in the local x and y directions of the pad plane respectively (i.e.\ parallel to the azimuthal direction and to the beam direction). These distributions are shown in Fig.~\ref{Fig:RichRecoRes}.
The TPC and RICH selection criteria are summarized in Table~\ref{Tab:rich_cuts}.\\

\begin{center}
\begin{figure}[ht!]
 \includegraphics[width=0.48\textwidth]{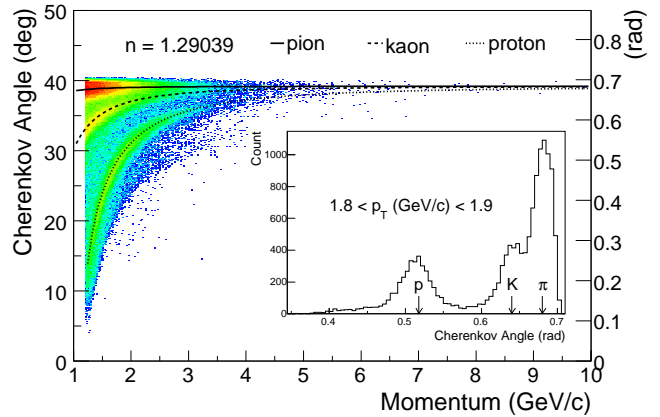} 
\caption{(Color online) Cherenkov angle vs. momentum (momentum obtained with the TPC). The inset shows the projection of the Cherenkov angle for a small $p_T$ interval.}
\label{Fig:RichSpectra}
\end{figure}
\end{center}

\begin{center}
\begin{figure}[ht!]
 \includegraphics[width=0.48\textwidth]{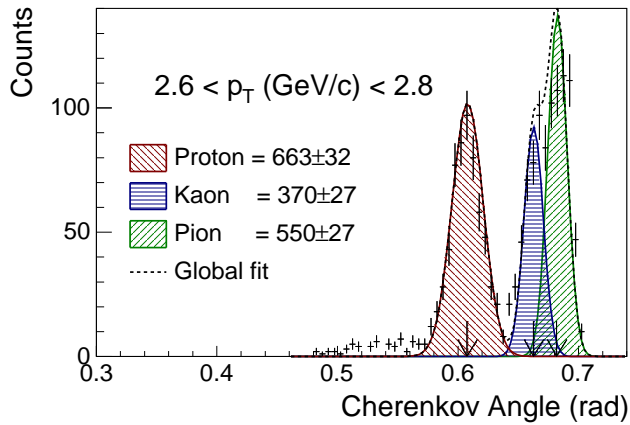} 
\caption{(Color online) Extracted Cherenkov angle distribution for particles with
a positive charge and $2.6<p_T<2.8~\mathrm{GeV}/c$. This $p_T$ range is close to the limit where mesons cannot be identified using the integrated fit technique. The
data used are those for the most central collisions. Using a constraint on the total number of counts gives a $\chi2/NDF=3.8$.}
\label{Fig:RichFit}
\end{figure}
\end{center}

The Cherenkov angle versus momentum for reconstructed charged particles is presented in Fig.~\ref{Fig:RichSpectra}.  Lines correspond to the expected angles for pions, kaons and protons for an index of refraction of $n=1.29039$~\cite{Ref:HmpidAlice}. By binning in transverse momentum, histograms such as the one shown in the insert of Fig.~\ref{Fig:RichSpectra} enable particle identification with the RICH as discussed below. 
Identification of mesons is possible up to $p_T=3~\mathrm{GeV}/c$ and up to $p_T=4.5~\mathrm{GeV}/c$ for protons and anti-protons once the cuts in Table~\ref{Tab:rich_cuts} are applied.
The momentum of each particle at the primary vertex is different from the momentum at the RICH radiator. In order to perform a three-Gaussian fit to the Cherenkov angle distribution in each $p_T$ interval and extract the raw yields, the momentum at the radiator level is needed. Therefore, the mean energy loss in the TPC and its outer field cage for the three different species (pion, kaon, and proton), as a function of momentum, is determined via Monte Carlo simulation.
After full propagation with GEANT and reconstruction, Gaussian fit parameters for the Monte Carlo spectra are compared to those extracted from real data.\\
We find very good agreement at low transverse momenta ($p_T<2~\mathrm{GeV}/c$), i.e.\ where the three peaks are sufficiently separated for unambiguous determination of the mean and width of each Gaussian fit to the real data.
With increasing $p_T$, the Gaussians start merging, and several fit strategies are used in order to extract the raw yields: i) require that the three-Gaussian integral equals the total number of projected tracks; ii) use the means and/or the widths  extracted from simulation to add more contraints; iii) when kaons and pions are difficult to disentangle (for $p_T>2.6~\mathrm{GeV}/c$), fix the kaon yields with the measured $K^{0}_{S}$ yields, presented later in this paper, as a unique constraint for obtaining the raw yields of protons, anti-protons and pions.
An example of a fit where the means of the Gaussians are fixed using simulation parameters is shown in Fig.~\ref{Fig:RichFit}. 
The arrows indicate the positions of the Gaussian means of the fits to the simulation data.
The errors shown in Fig.~\ref{Fig:RichFit} are quadratic sums of statistical errors and systematic uncertainties due to the fitting.
The overall systematic uncertainties for the extracted raw yields are obtained by changing the fitting techniques, varying the number of free parameters (leaving free and fixing means, widths, or requiring that the overall integral is equal to the number of counts), as well as varying the cuts of Table~\ref{Tab:rich_cuts}.

\subsection{Topological identification of $\Lambda$ and $K^{0}_{S}$ with the TPC.}
The strange hadrons, $\Lambda$ and $K^{0}_{S}$ (so called ``V0s''), are reconstructed in the Time Projection Chamber at mid-rapidity ($|y| <$ 1) via  measurements of their charged decay products (daughters) from weak decays: $\Lambda(\overline{\Lambda}) \rightarrow p(\overline{p})+\pi^{-}(\pi^{+})$ (64$\%$) and  $K^{0}_{S} \rightarrow \pi^{+}+\pi^{-}$ (69$\%$).  The strange hadrons are reconstructed by pairing tracks of opposite sign and a small distance of closest approach (DCA), while ensuring that the momentum vector of the assumed neutral parent is on a line through the interaction vertex.  However, due to the large particle multiplicities in  relativistic heavy-ion collisions, particle spectra produced in this way suffer from very large combinatorial backgrounds, especially at low $p_T$, owing to random crossings of uncorrelated tracks predominantly close to the interaction vertex.  In order to minimize this background, geometrical selections (i.e., cuts) are applied to the data which reduce the background while keeping as much of the signal as possible.  These selections include imposing a minimum decay length for the V0, which removes all crossings of primary tracks at the interaction vertex which would appear as V0s with zero decay length, as well as cuts on the minimum DCA of the daughter particles to the interaction vertex.  These selections are effective as the momentum vectors from true daughters do not lie along a line through the interaction vertex owing to the momentum kick given to these particles in the decay (larger for $K^{0}_{S}$ compared to $\Lambda$ and $\overline{\Lambda}$ due to the differences in mass between the parent and sum of the daughter particles).  However, this DCA sensitivity is found to be strongly dependent upon the $p_T$ of the parent particle, hence different selections were utilized as a function of the parent $p_T$.  The values used for these selections for the three different particles are given in Table~\ref{Tab:Cuts}.
\begin{table*}
\begin{ruledtabular}
\begin{tabular}{cccc}
Geometrical Selections & $\Lambda$ & $\overline{\Lambda}$ & $K^{0}_{S}$\\
\hline
DCA V0 to primary vertex &$<$ 0.5 cm & $<$ 0.5 cm & $<$ 0.75 cm  \\
DCA neg daughter to primary vertex\footnotemark[1] &$>$ 2.5 cm : $>$ 4.0 cm & $>$ 1.5 cm : $>$ 0.0 cm & $>$ 3.5 cm : $>$ 2.0 cm : $>$ 0.6 cm \\
DCA pos daughter to primary vertex\footnotemark[1] & $>$ 1.5 cm : $>$ 0.0 cm & $>$ 2.5 cm : $>$ 4.0 cm & $>$ 3.5 cm : $>$ 2.0 cm : $>$ 0.6 cm\\
DCA daughters at secondary vertex & $<$ 0.8 cm & $<$ 0.8 cm & $<$ 0.8 cm\\
Decay length & $>$ 6 cm & $>$ 6 cm & $>$ 5 cm \\
\end{tabular}
$^{a}$This cut is strongly dependent upon the transverse momentum ($p_{T}$).  For the $\Lambda$ and $\overline{\Lambda}$, two different $p_{T}$ ranges were used, $p_{T} <$ 3 GeV/$c$, $p_{T} >$ 3 GeV/$c$.  For the $K^0_{S}$, three $p_{T}$ ranges were used, $p_{T} <$ 1.0 GeV/$c$, 1.0 $< p_{T} <$ 3.5 GeV/$c$, $p_{T} >$ 3.5 GeV/$c$.
\caption{The values of the geometrical selection variables used to reconstruct the strange hadrons.}
\label{Tab:Cuts}
\end{ruledtabular}
\end{table*}

\begin{center}
\begin{figure}[ht!]
 \includegraphics[width=0.48\textwidth]{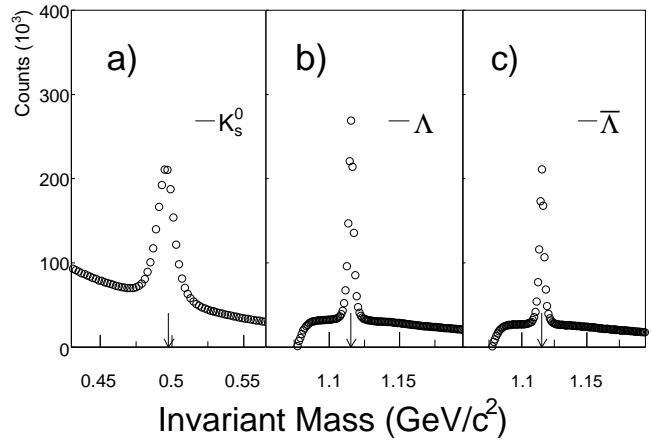} 
\caption{The relative raw mass spectra for the $K^{0}_{S}$, $\Lambda$ and $\overline{\Lambda}$ candidates, respectively, within a given low transverse momentum bin.}
\label{Fig:Masses}
\end{figure}
\end{center}

\begin{center}
\begin{figure}[ht!]
 \includegraphics[width=0.48\textwidth,height=0.38\textwidth]{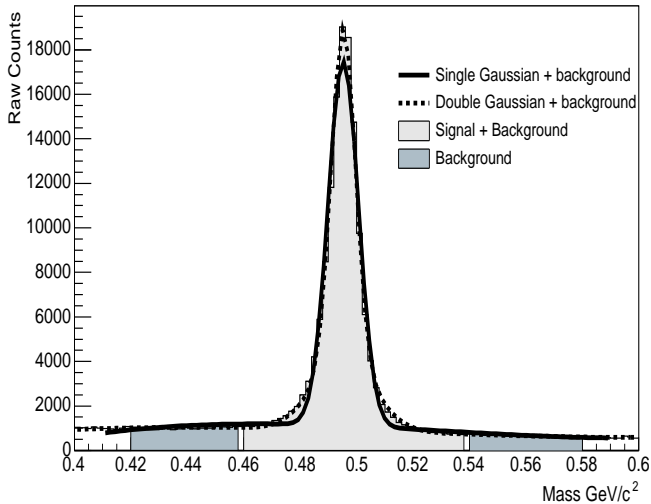} 
\caption{The raw mass spectrum for the $K^{0}_{S}$ for a given transverse momentum bin, with three different signal extraction methods indicated.}
\label{Fig:FitLambda}
\end{figure}
\end{center}

The raw mass spectra for the different hadrons  are shown in Fig.~\ref{Fig:Masses} for a typical low transverse momentum bin where the backgrounds are the most significant.
In order to extract  raw yields, two different methods are employed.  The first uses a fit function to describe both the signal (single-Gaussian or double-Gaussian) and the background on either side of the peak (third-order polynomial).  For the example shown in Fig.~\ref{Fig:FitLambda},  the single Gaussian method fails to fit the signal, both in the tail regions of the distribution and the magnitude of the peak; however, for the example shown, the double Gaussian method reproduces the data well with a small  $\chi^{2}$ for the fit and does so in the majority of the other cases.  Occasionally, however, this method fails, and the fit does not converge.  To avoid such cases, another method of extracting the raw yields is used.  This method involves the summing of the entries of relevant bins in the histogram as shown in Fig.~\ref{Fig:FitLambda}, and is referred to as `bin-counting'~\cite{Ref:STAR_130GeV}.  Three regions of the histogram are identified: sidebands on each side of the mass peak (each of width $15~\mathrm{MeV}$ taken to represent the sum of the background under the peak) and the signal area itself (in this case, $\pm15~\mathrm{MeV}$ of the mass of the $K^{0}_{S}$) which contains both signal and background.  The raw yield is the difference between the signal area and the background area. This method works best if the shape of the background is linear over the full range.  If the background is small in comparison to the signal, then any small deviations from linearity of the background have  negligible effect on the extracted yield.
\subsection{Correction Factors}
\begin{center}
\begin{figure}[ht!]
 \includegraphics[width=0.48\textwidth]{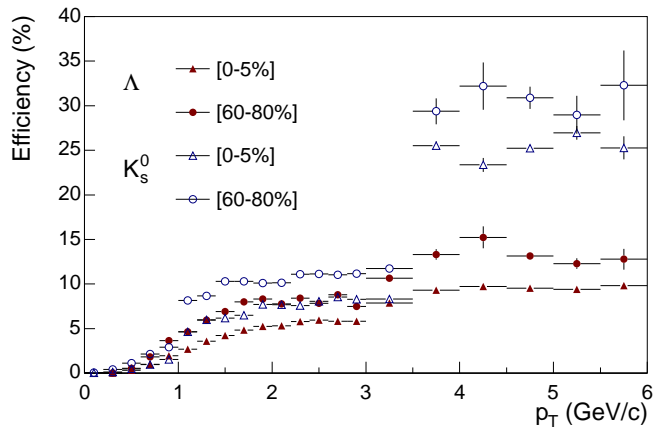} 
\caption{(Color online) Efficiencies as a function of $p_T$ and centrality for the $K^{0}_{S}$ and $\Lambda$.  The three ranges in the efficiency are due to  changing of geometrical reconstruction cuts given in Table~\ref{Tab:Cuts}, chosen for optimizing the signal-to-noise ratios and maximizing the statistics at high $p_T$.}
\label{Fig:CorrFact}
\end{figure}
\end{center}

\begin{center}
\begin{figure}[ht!]
 \includegraphics[width=0.48\textwidth]{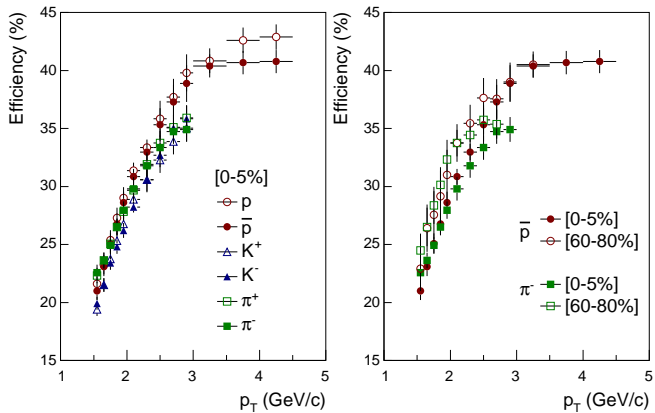} 
\caption{(Color online) Charged particle reconstruction and identification efficiencies as a function of $p_T$ (left panel). Collision centrality effects, mainly due to TPC track reconstruction, are shown for the anti-proton and the negative pion for the extreme centrality bins used (right panel).}
\label{Fig:RichEffi}
\end{figure}
\end{center}

After the raw yields are extracted as a function of $p_T$, they must be corrected for analysis inefficiencies in order to obtain final spectra.
These inefficiencies are calculated by embedding Monte Carlo (MC) tracks into real data and determining the probability of finding these particles through data analysis.
This process is performed in STAR utilizing the GEANT code.
The MC particles are generated with a user-defined realistic distribution in transverse momentum and rapidity space.
As the real data have a finite interaction vertex distribution, care is taken to ensure that the generated particles have the same interaction vertex distribution as the real data.
GEANT propagates the particles through the material present in STAR, and the resultant ionization deposited in the TPC is then effectively drifted to the read-out planes using a detailed simulation response program which provides ADC values from the cathode pads that are in the same format as that of the real data.
The data from the simulation are then mixed with real data, and STAR's full tracking and reconstruction software is used on the mixed data to determine the efficiency for finding the MC particles.  As the MC information is recorded separately for the embedded particles, it is easy to determine the efficiency of the STAR analysis software for finding the embedded particles. The same geometrical and multiplicity cuts are used for the mixed data as are used for the real data.
The total efficiency, for each $p_T$ bin, is simply the ratio of the number of reconstructed MC particles to the number of generated particles.
As the majority of the background is dependent upon the track multiplicity in the event, it is found that the efficiency is also strongly dependent upon collision centrality.
Care is taken to ensure that when the mixing is done, the multiplicity of the mixed event is not increased by more than 5$\%$ of its original value.
This value is  chosen so that the efficiency is not compromised while maximizing the available statistics.
In addition to the TPC MC simulation, a RICH MC simulation is done.  Here, the ADC values corresponding to both the {\sc mip} and the photons generated as the particles pass through the radiator are generated and used in the efficiency calculation.  This is done during the same reconstruction pass as for the TPC embedding.
Examples of the centrality and $p_T$ dependence of the efficiencies are shown in Figs.~\ref{Fig:CorrFact} and~\ref{Fig:RichEffi}, which show efficiencies for the most central and most peripheral centrality classes used in the analysis presented in this paper.\\

\begin{center}
\begin{figure}[ht!]
 \includegraphics[width=0.48\textwidth]{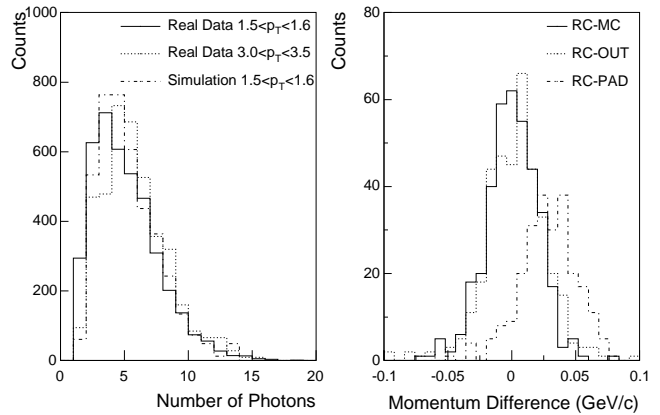} 
\caption{(left panel) A comparison of the number of photons  for different $p_T$ bins for protons. (right panel) The energy loss effects on momentum determination: simulated (MC), reconstructed at the primary vertex (RC), reconstructed at the outermost point of the TPC (OUT), as well as at the RICH pad level (PAD).}
\label{Fig:RichRecoPhoton}
\end{figure}
\end{center}

At relatively low transverse momenta a $10\%$ difference is observed among charged mesons because kaons can decay before reaching the RICH fiducial volume (left panel of Fig.~\ref{Fig:RichEffi}).
Within statistical errors no significant difference is observed between opposite charges of the same species, except for baryons, due to anti-proton absorption in the detector material.
Although not shown for clarity in the right panel of Fig.~\ref{Fig:RichEffi}, efficiencies for the intermediate centrality bins ([20--40$\%$] and [40--60$\%$]) sit between the extremes within statistical errors.
This dependence on centrality is not a feature of the RICH due to low occupancy, but rather comes from the track finding inefficiency in the TPC, as a track has to be found in the TPC before it can be extrapolated to the RICH.
The number of photo-electrons in a cluster for an identified Cherenkov ring is a function of the relativistic velocity and therefore of the momentum of the particle at the radiator level; it increases with $\beta$ and reaches up to $17\pm2$ when $\beta \simeq 1$~\cite{Ref:RICHTest1,Ref:RICHTest2,Ref:RICHTest3,Ref:RICHTest4}. The left panel of Fig.~\ref{Fig:RichRecoPhoton} shows distributions of the number of photons for the limiting cases of protons which have a significantly lower velocity for a fixed momemtum bin. The lower limit for the mean number of photons (for $1.5<p_T<1.6~(\mathrm{GeV}/c))$ is approximately 5 for this species and is correctly reproduced by the simulation (dashed line).
Due to energy loss along the path from the interaction vertex to the RICH pad plane, significant differences can be seen in the momentum of the same particle at different points along its path. The right panel of Fig.~\ref{Fig:RichRecoPhoton} presents distributions of the differences of these momenta for simulated protons at low momentum where the differences can be relatively significant: no difference is seen between the simulated momentum (MC) and the one reconstructed at the primary vertex (RC) within momentum determination uncertainties.
Though a small shift is distinguishable for the momentum at the end of the TPC (OUT), the most significant one is at the RICH pad level (PAD) once the track has travelled through the TPC, the outer field cage and the Cherenkov radiator. The mean momentum shift resulting from the overall energy loss is a function of the track momentum. It is obtained from simulation and taken into account to determine the corrected mean Cherenkov angle for each $p_T$ bin.\\
\section{Results}

\begin{center}
\begin{figure}[ht!]
 \includegraphics[width=0.48\textwidth]{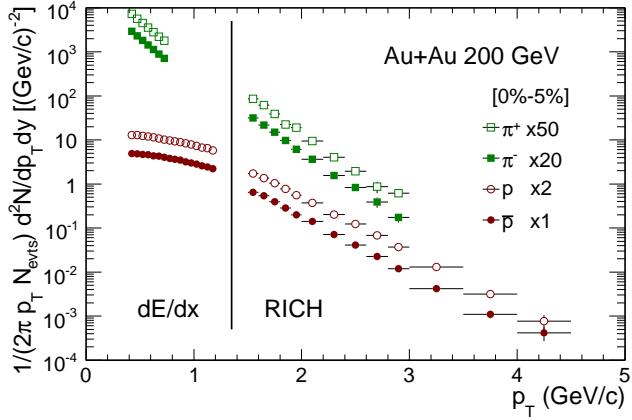}
 \caption{(Color online) Identified charged particle spectra from analysis of the RICH data (right of vertical line); to the left of the vertical line are identified charged particle spectra determined using energy loss ($dE/dx$) in the TPC~\cite{Ref:OlgaPRL}.}
 \label{Fig:RICHSpectra}
\end{figure}
\end{center}

\begin{center}
\begin{figure}[ht!]
 \includegraphics[width=0.48\textwidth]{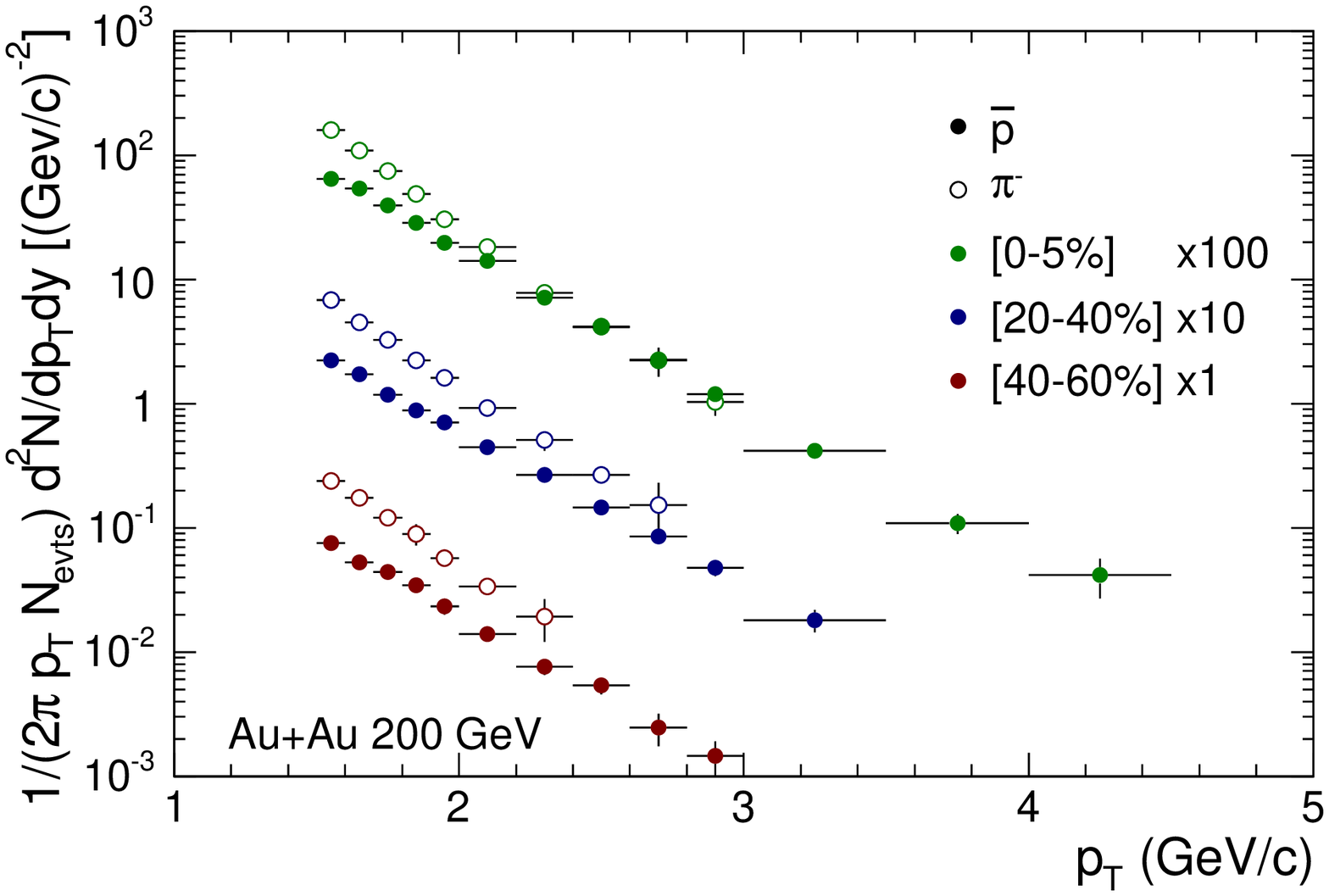}
 \caption{(Color online) The $\overline{p}$ (full symbols) and $\pi^{-}$ (open symbols) spectra as a function of both centrality and $p_{T}$.}
 \label{Fig:RICHSpectra2}
\end{figure}
\end{center}

\begin{center}
\begin{figure}[ht!]
 \includegraphics[width=0.48\textwidth]{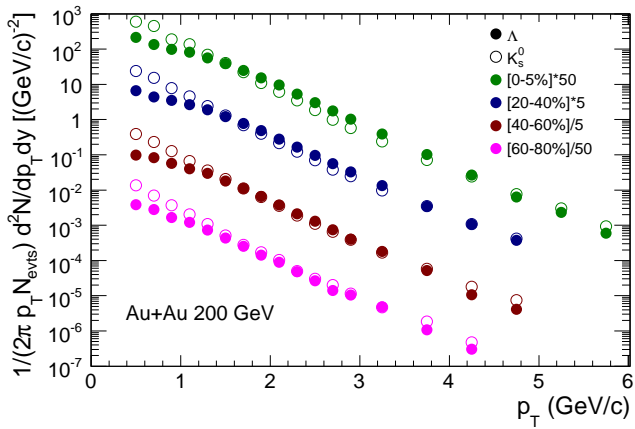} 
\caption{(Color online) The $\Lambda$ (full symbols) and K$^0_S$ (open symbols) spectra as a function of both centrality and $p_T$.}
\label{Fig:StrangeSpectra}
\end{figure}
\end{center}

Once all efficiencies and correction factors are calculated (both as a function of $p_T$ and centrality), they are applied to the raw data\textcolor{blue}{,} and final spectra are obtained as illustrated in Figs.~\ref{Fig:RICHSpectra}, \ref{Fig:RICHSpectra2} and~\ref{Fig:StrangeSpectra}. 

Figure \ref{Fig:RICHSpectra} shows a comparison between the non-strange identified spectra from the RICH analysis and those determined from energy loss ($dE/dx$) in the gas volume of the TPC for the most central bin~\cite{Ref:OlgaPRL}.  The extra reach of identified particles obtained using the RICH detector is significant.  Figure \ref{Fig:RICHSpectra2} shows the $\overline{p}$ and $\pi^{-}$ spectra from the RICH for  three of the centrality classes studied, while Figure \ref{Fig:StrangeSpectra} shows  the strange particle spectra for all four centralities used.

The maximum extent in $p_T$ for the $\Lambda$ and $K^{0}_{S}$ is not restricted by the identification method, but rather is limited by statistics only.  Therefore, the reach in $p_T$ for the [0--5$\%$] centrality bin extends further than for the other centrality bins as it includes data taken with the most central trigger, mixed with the most central portion of the data taken with the minimum bias trigger.

The results shown in Fig.~\ref{Fig:StrangeSpectra} take into account both statistical and systematic errors (added in quadrature); their values are smaller than the symbol sizes.  For both particles, the systematic errors are dominated by signal extraction uncertainties, which vary from 5(20)$\%$ at very low $p_T$ for $K^{0}_{S}$($\Lambda$), to 10(10)$\%$ at high $p_T$.  The other sources of systematic error include effects due to magnetic field polarity, effects due to differences between TPC East data and TPC West data, as well as effects due to the choice of geometrical cuts. We estimate that these effects contribute an additional 5$\%$ to the uncertainty for both the $K^{0}_{S}$ and $\Lambda$.

Neither the anti-proton nor $\Lambda$ hyperon spectra presented in this paper are corrected for feed-down from weak decays.  The dominant channels were studied for each case and found to be 15-20$\%$ for the anti-protons,  10$\%$  for the $\Lambda$, and independent of transverse momentum.

\begin{center}
\begin{figure}[ht!]
 \includegraphics[width=0.48\textwidth]{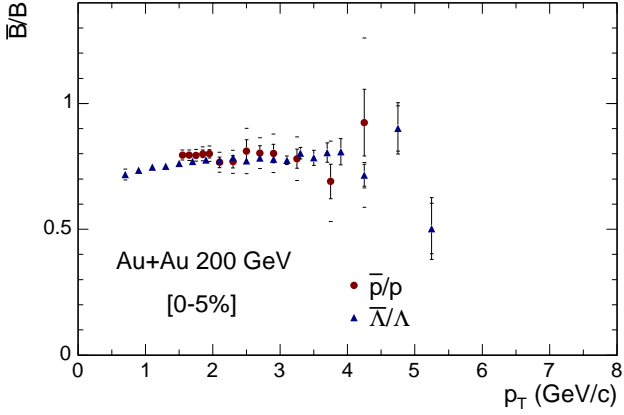} 
\caption{(Color online) The $\overline{p}/p$ (circles) and $\overline{\Lambda}/\Lambda$ (triangles) ratios as a function of $p_T$.  The statistical errors are indicated by the error bars, the systematic errors are indicated by the hash marks. }
\label{Fig:BBarB}
\end{figure}
\end{center}

The $\overline{p}/p$ and $\overline{\Lambda}/\Lambda$ ratios are plotted in Fig.~\ref{Fig:BBarB} as a function of $p_T$ for the [0--5$\%$] centrality bin. Both statistical and systematic errors  are shown.
For both cases, raw particle yields are used because of the charge symmetric nature of both the applied cuts for particle reconstruction and the tracks.  
The ratios are corrected for absorption in the detector material using a GEANT simulation;  the largest absorption correction occurs for the anti-baryon.  
It is expected that at large $p_T$, where particle production is dominated by jet fragmentation, the anti-baryon/baryon ratio ($\overline{B}/B$) will start to decrease.  
This is because the leading baryon from a quark jet is more likely to be a particle rather than an anti-particle, whereas this asymmetry does not occur in a gluon jet. 
For the most central collisions, this trend is not observed in either the $\overline{p}/p$ ratio out to $p_T=4.5~\mathrm{GeV}/c$ or the $\overline{\Lambda}/\Lambda$ ratio, out to $5.5~\mathrm{GeV}/c$.  
The ratios for the other centrality bins are not plotted for clarity. However,
as a function of centrality, neither ratio shows a decrease though the $p_T$ coverage is less, due to lack of statistics.  The integrated value of the ratio is constant for all centralities.

The baryon to meson ratios ($B/M$) can also be studied with this data-set to investigate further the relative baryon enhancement first reported in the $130~\mathrm{GeV}$ data.  
That work showed that the $\Lambda/h^-$ ratio approached unity in the intermediate $p_T$ region~\cite{Ref:STAR_130GeV}, as did the $p/\pi$ ratio~\cite{Ref:PHENIX_130GeV}.  
This is contrary to what is observed for elementary particle collisions where the ratio does not exceed 0.2 in the corresponding momentum range for both quark and gluon jets~\cite{Ref:DELPHI_Ratios}.  
This is as expected for jet fragmentation.

\begin{center}
\begin{figure}[ht!]
 \includegraphics[width=0.48\textwidth]{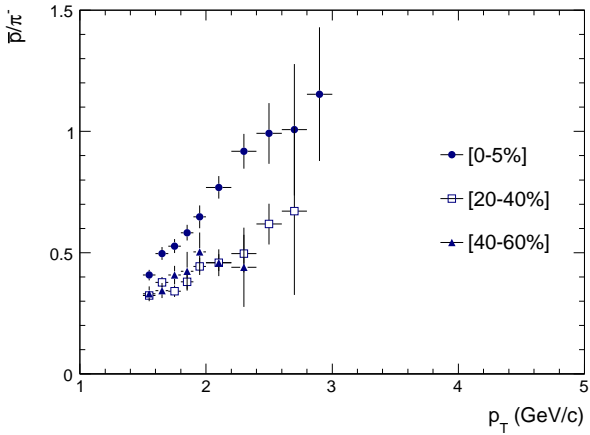} 
\caption{The $\overline{p}/\pi^{-}$ ratio as a function of centrality and $p_T$ for Au+Au collisions at $\sqrt{s_{NN}}=200~\mathrm{GeV}$.}
\label{Fig:PbarPi}
\end{figure}
\end{center}

Figure~\ref{Fig:PbarPi} shows the $\overline{p}/\pi^{-}$ ratios, again for the 5$\%$ most central collisions, as well as for both the [20--40$\%$] and [40--60$\%$] centrality classes.  The $p_T$ range is limited to $3~\mathrm{GeV}/c$  due to the need for unambiguous identification of the $\pi^{-}$ and by statistics for the more peripheral samples.  The ratios clearly increase with increasing $p_T$ and also exhibit  centrality dependence; 
for a given $p_T$, the magnitude is larger for the more central collisions. However, for the $p_T$ range studied here,  there is no evidence for the ratio turning over or decreasing to the small values measured in jet production~\cite{Ref:DELPHI_Ratios}. It should occur, however, for high enough momenta when fragmentation is the dominant process.  We note here that this ratio is consistent with the ratio presented by the PHENIX collaboration once the differences in centrality and treatment of feed down from weak decays are taken into account~\cite{Ref:PHENIX_200GeV}. 

\begin{center}
\begin{figure}[ht!]
\includegraphics[width=0.48\textwidth]{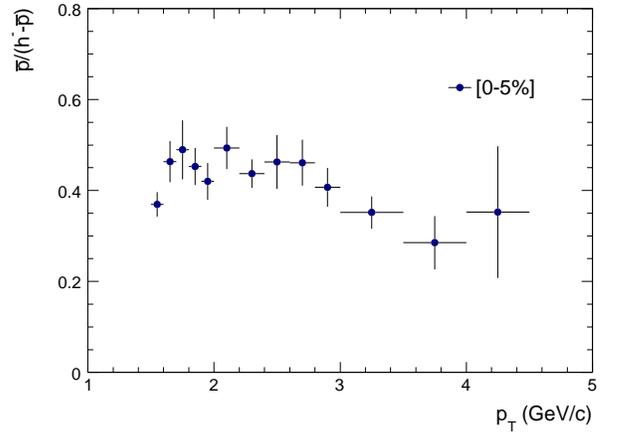} 
\caption{The $\overline{p}/(h^- - \overline{p})$ ratio as a function of $p_T$ for Au+Au collisions at $\sqrt{s_{NN}}=200~\mathrm{GeV}$.}
\label{Fig:PbarHmin}
\end{figure}
\end{center}

As the identification of $p$ and $\overline{p}$ using the STAR RICH detector extends well beyond that of the $\pi^-$, it is still possible to measure an approximation of the $B/M$ ratio to higher values of $p_T$ by plotting the $\overline{p}/(h^- - \overline{p})$ ratio as a function of $p_T$, where the value $h^-$ represents the sum of non-identified negatively charged hadrons ($\pi^-, K^-$ and $\overline{p}$).  As the identification of the particles is not required for the $h^-$ data, it is possible to use the TPC data which has a larger acceptance than the RICH detector and hence the reach in $p_T$ for the $\overline{p}/(h^- - \overline{p})$ ratio is dominated by the reach in $p_T$ of the identified $\overline{p}$.  The $\overline{p}$ spectrum has been subtracted in the denominator because it is part of the measured $h^{-}$ spectrum.
This ratio is plotted in Fig.~\ref{Fig:PbarHmin} which shows that the reach in $p_T$ extends to $4.5~\mathrm{GeV}/c$.
With the addition of the kaons in the denominator, the ratio shows an initial shape which is different from that of the $\overline{p}/\pi^-$ ratio, but the region of importance is for $p_T>3~\mathrm{GeV}/c$ where the trend of the data is for a slight decrease in the ratio.

\begin{center}
\begin{figure}[ht!]
 \includegraphics[width=0.48\textwidth,height=0.38\textwidth]{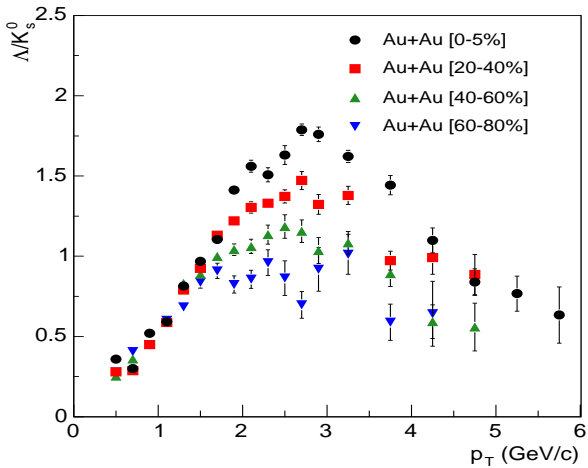} 
\caption{(Color online) The $\Lambda/K^{0}_{S}$ ratio as a function of centrality and $p_T$ for Au+Au collisions at $\sqrt{s_{NN}}=200~\mathrm{GeV}$.}
\label{Fig:LamK0s}
\end{figure}
\end{center}

In order to extend the reach in $p_T$ even further, we  use the strange particles which can be identified out to $6~\mathrm{GeV}/c$ in the most central bin (as illustrated in Fig.~\ref{Fig:StrangeSpectra}).  The $\Lambda/K^{0}_{S}$ ratio, a measure of the baryon/meson ratio in the strangeness sector, is plotted in Fig.~\ref{Fig:LamK0s} as a function of $p_T$ for the four  centrality classes studied.
The quoted error bars are statistical only as the calculated systematic errors for the $\Lambda$ and $K^{0}_{S}$ are correlated to an unknown degree, resulting in smaller systematic errors.
With the increased  $p_T$ range for the $B/M$ ratio, a definite turnover for strange particles is observed  at $p_T \approx~ 3~\mathrm{GeV}/c$ for all centrality classes.  For $p_T > 3~\mathrm{GeV}/c$  for each centrality bin the ratio decreases. 
At low transverse momenta  the ratios appear to be independent of centrality (up to 
$p_T \sim 1.5~\mathrm{GeV}/c$).  Beyond $p_T \sim 1.5~\mathrm{GeV}/c$, at each $p_T$, the ratios increase with increasing centrality.

The $\Lambda/K^{0}_{S}$ ratio is greater than unity 
for the [0-5$\%$] and [20-40$\%$] data in the  $p_T$ range $1.5<p_T<5~\mathrm{GeV}/c$; in all cases it is larger than the $B/M$ ratio measured for elementary collisions~\cite{Ref:DELPHI_Ratios}. 

For the first time, it is shown that  the $\Lambda$/$K^{0}_{S}$ ratio (and hence the $B/M$ ratio) does not keep rising with increasing $p_T$, but rather decreases for the region $\sim 3 < p_T < 5~\mathrm{GeV}/c$.

The data suggest that at  $p_T~ 5~\mathrm{GeV}/c$
the ratios for each centrality bin are roughly equal.  This observation then suggests that the region
$\sim~ 2 <p_T < 5~\mathrm{GeV}/c$ is the intermediate $p_T$ region to use to test the ideas discussed in the introduction.

In the following section, we compare a number of theoretical models to the data and discuss their applicability.   Each model uses a different method to describe particle production in the intermediate $p_{T}$ region.

\section{Discussion}

\begin{center}
\begin{figure}[ht!]
 \includegraphics[width=0.48\textwidth]{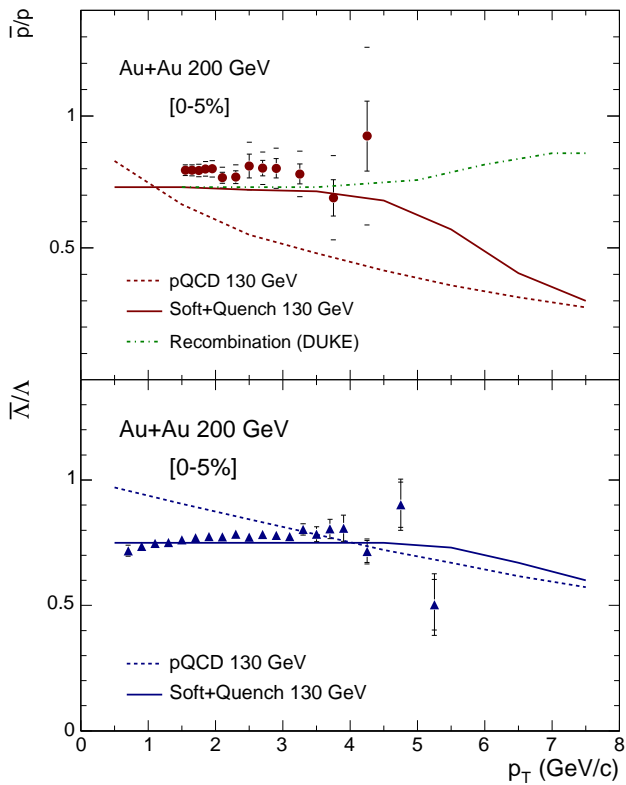} 
\caption{(Color online) The $\overline{p}/p$ (upper plot) and $\overline{\Lambda}/\Lambda$ (lower plot) ratios as a function of $p_T$, together with various model predictions (see text for full description).}
\label{Fig:BBarB_models}
\end{figure}
\end{center}

The $\overline{p}/p$ and $\overline{\Lambda}/\Lambda$ ratios, presented earlier, are plotted in Fig.~\ref{Fig:BBarB_models} as  functions of $p_T$ for the [0--5$\%$] centrality bin,  together with predictions of various theoretical models. 
The main feature of 
both ratios is that they are essentially constant as a function of $p_T$. 
This observation contradicts predictions from pQCD inspired models which show a decrease in the ratio over all $p_T$~\cite{Ref:Wang_PRC,Ref:Vitev_QM}.  
The prediction from one of these models (\cite{Ref:Vitev_QM}) is shown in Fig.~\ref{Fig:BBarB_models}, though the prediction is for Au+Au at $\sqrt{s_{NN}}=130~\mathrm{GeV}$.  
The reason for the  decrease is that in a quark jet, the leading baryon is more likely to be a particle than an anti-particle, whereas in a gluon jet, there is no such asymmetry.  
Although the pQCD prediction  extends to below $p_T=1~\mathrm{GeV}/c$, it is not expected to be realistic in the low $p_T$ region as particle production is not dominated by jets, but rather is governed by soft particle production and hydrodynamics~\cite{Ref:HydroSTARData,Ref:HydroModels}.  
The `Soft+Quench' model, however, is in good agreement with the data and predicts $\overline{p}/p$ and $\overline{\Lambda}/\Lambda$ ratios which are flat up
to about $p_T=5~\mathrm{GeV}/c$.  
This model links soft particle production at low $p_T$ (where particle production is assumed to be proportional to $e^{-p_{T}/T}$, $T$ being the inverse slope parameter) with the higher $p_T$ region dominated by a leading order pQCD calculation incorporating gluonic baryon junctions~\cite{Ref:Vitev_QM}.  
The model also includes initial state multiple scattering (related to the ``Cronin effect"~\cite{Ref:Cronin}),
 nuclear shadowing and partonic energy loss due to gluon bremsstrahlung as calculated in the Gyulassy-Levai-Vitev (GLV) formalism~\cite{Ref:GLV}.  
The third model prediction in Fig.~\ref{Fig:BBarB_models} is from a recombination model ($DUKE$) which reproduces the $\overline{p}/p$ ratio well in the range where there is data~\cite{Ref:Fries_PRC}.
The recombination model itself is again governed by soft processes at low $p_T$ and at high $p_T$ by jet fragmentation in addition to jet quenching mechanisms; the predicted ratio starts to decrease for $p_T > 8~\mathrm{GeV}/c$.  
The model assumes that recombination processes (where two quarks combine into mesons and three combine to form baryons) are dominant at low $p_T$ where the partonic spectrum is exponential in shape, while fragmentation becomes dominant in the high $p_T$ region when the parton spectrum exhibits a power-law behavior.  In this implementation, jet quenching mechanisms are also applied to the high $p_T$ region.
Therefore, in this model, the intermediate $p_T$ region ($2<p_T<5~\mathrm{GeV}/c$) is dominated by recombination.

\begin{center}
\begin{figure*}[t!]
 \includegraphics[width=0.68\textwidth]{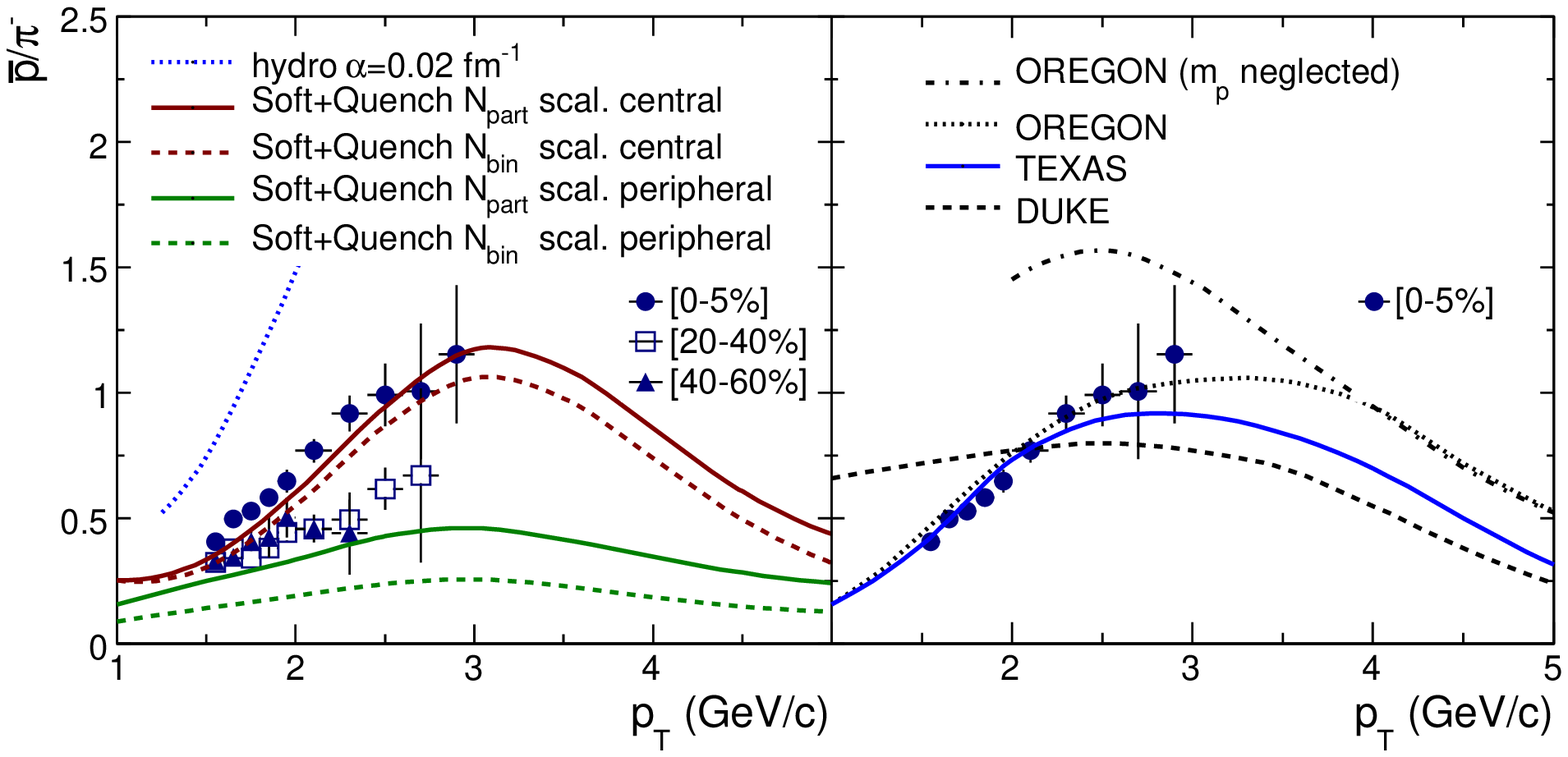} 
\caption{(Color online) The $\overline{p}/\pi^{-}$ ratio for the [0--5$\%$], [20--40$\%$] and [40--60$\%$] most central collisions together with different model predictions for the most central bin (together with a prediction for a peripheral bin for the `Soft+Quench' model).  For the case of the {\it DUKE} model, the prediction is for $\overline{p}/\pi^{0}$.}
\label{Fig:PbarPiModel}
\end{figure*}
\end{center}

Turning to the $B/M$ ratios, Fig.~\ref{Fig:PbarPiModel} shows  comparisons between the $\overline{p}/\pi^{-}$ ratio as a function of centrality and predictions from a variety of models.  
Besides results from the models described above, predictions are also shown for a hydrodynamical model with a soft equation of state~\cite{Ref:Kolb}, and from  two additional recombination models~\cite{Ref:PrcCoalHwa,Ref:PrcCoalGKL}.  

Although hydrodynamical models have been shown to work well at low $p_T$~\cite{Ref:Kolb}, at the higher $p_T$ (see left panel of Fig.~\ref{Fig:PbarPiModel}), they cannot reproduce the data.  In this model, $\alpha$ is proportional to an initial transverse seed velocity.  
The major discrepancy can be attributed to the predicted $\pi^-$ spectrum which deviates from the measured data at  $p_T \approx 1~\mathrm{GeV}/c$. 
Conversely, the `Soft+Quench' model provides a qualitatively good description of the data, matching the observed rise in the data as well as the larger magnitude of the ratio for the more central data-set~\cite{Ref:Vitev200}.
For this model, there are predictions for two different geometrical scalings. N$_{part}$ scaling refers to scaling with the number of participants, while N$_{bin}$ scaling is with respect to the number of binary collisions.  The largest difference between these scalings occurs in the more peripheral bin where the N$_{part}$ scaling gives better agreement with the data.

The `Soft+Quench' model predicts the ratio's behavior to be due mainly to  two effects.  Firstly, pion production in central collisions is found experimentally to be lower than that in $p+p$ collisions when scaled by the relevant number of binary collisions~\cite{Ref:PHENIX_200GeV}.  This is attributed to quenching of jets due to high $p_T$ particles interacting with a dense gluonic medium.  Secondly, baryon production is dominated at low to intermediate $p_T$ by non-perturbative baryon junctions.  The junctions themselves are not found to be affected by the jet quenching, and hence the baryons scale with binary collisions from $p+p$ data as expected.  This scaling leads to an enhanced baryon/meson ratio at intermediate $p_T$.  It only decreases at higher momenta when perturbative processes become relevant for baryon production.  The dependence of the ratio on centrality is naturally explained by the lack of jet quenching in the more peripheral collisions owing to the smaller, less dense medium that is created.

On the right panel of Fig.~\ref{Fig:PbarPiModel}, three more model predictions are compared to the experimental results.
These models are similar in that they all subscribe to a recombination (or coalescence) mechanism of dressed (i.e.\ massive) quarks, but they have different implementations.  
In the following discussion, all references to partons pertaining to the coalescence models will  refer to massive quarks.  

In each case, the models attribute particle production to the recombination mechanism when the parton transverse momentum spectra are exponential.  At $\sim p_T > 5~\mathrm{GeV}/c$, where it is believed that the parton distribution is governed by a power-law, fragmentation dominates.  Particle production at low $p_T$ ($<1~\mathrm{GeV}/c$) is governed by hydrodynamics.

There are two methods used to evaluate the recombination integral.  
The first is an analytical solution, though assumptions are made which cause a break down at low $p_T$.  
The second  uses a Monte Carlo technique to evaluate the integral, negating the need for any approximations.  
Different approaches for the two methods are taken as described below.

The first recombination model ({\it DUKE}~\cite{Ref:Fries_PRC} -- discussed earlier in the ${\overline{p}}/{p}$ section) uses the approximation technique to do the recombination integral and allows for the recombination of uncorrelated thermal, valence quarks (i.e., quarks with an effective mass) only.  
The prediction is shown by the dashed line (right panel, Fig.~\ref{Fig:PbarPiModel}).  
For the hard scattering part of the spectrum, the partons are taken from a mini-jet calculation which includes energy loss~\cite{Ref:Fries_PRC,Ref:Fries_PRL}.

The second recombination model result shown in the right panel of Fig.~\ref{Fig:PbarPiModel} by the solid line ({\it TEXAS} ~\cite{Ref:PrcCoalGKL}) uses the Monte Carlo method to do the integral.  As well as allowing the recombination of thermal quarks, it also allows the recombination of thermal quarks with co-moving quarks coming from mini-jets~\cite{Ref:PrcCoalGKL}.  This has the effect of increasing the ratio in the intermediate $p_T$ range.

\begin{center}
\begin{figure*}[ht!]
 \includegraphics[width=0.68\textwidth]{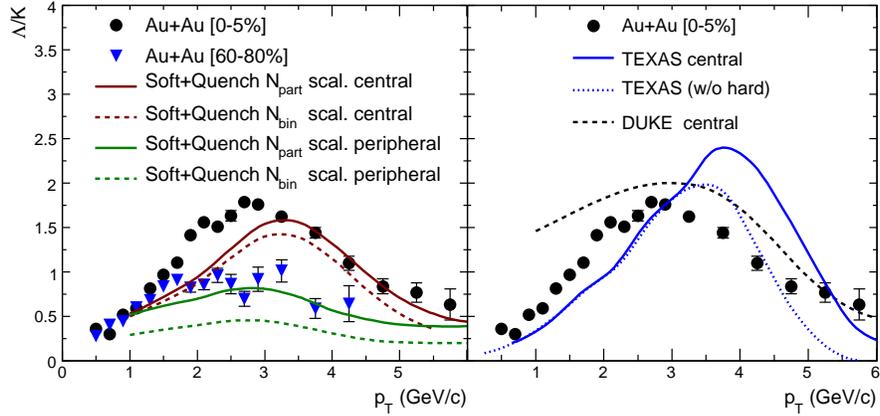} 
\caption{(Color online) The $\Lambda/K^{0}_{S}$ ratio as a function of centrality and $p_T$ for Au+Au  collisions at $\sqrt{s_{NN}}=200~\mathrm{GeV}$, together with various model predictions.}
\label{Fig:LamK0sModel}
\end{figure*}
\end{center}

The third recombination model ({\it OREGON} ~\cite{Ref:PrcCoalHwa}) again allows for recombination of thermal quarks but differs again in how  the higher $p_T$ regions are treated.  
Fragmentation functions are not used in the intermediate $p_{T}$ region, but instead, mini-jets are allowed to form  showers which then recombine; this is the original use of recombination models~\cite{Ref:HwaOriginal}.  
The shower partons are then allowed to recombine with the thermal quarks.
Results from two versions of this model are shown in Fig.~\ref{Fig:PbarPiModel}; the dashed-dotted line represents the original prediction where the mass of the proton is neglected, and hence below  $p_T~ \sim ~ 2~\mathrm{GeV}/c$, its assumptions become tenuous.  The small dotted line represents the result of a later calculation where transverse mass is used  instead of the proton mass~\cite{Ref:PrcCoalHwa2}.
The original calculation reproduces the proton spectra well, whereas the later one is in agreement with the $\overline{p}/\pi^{-}$ data.
It is evident from Fig.~\ref{Fig:PbarPiModel} that the recombination models, though all showing a $\overline{p}/\pi^{-}$ ratio which increases with $p_T$
up to about $3~\mathrm{GeV}/c$, have different behaviors at low $p_T$.
However, in common with the `Soft+Quench' model, they show a turnover to a falling ratio at $p_T~ \sim ~3~\mathrm{GeV}/c$, exactly where the data stop.
Therefore, we use the ratio of strange baryons ($\Lambda$) to strange mesons ($K^{0}_{S}$) to investigate this turnover region.
This comparison is made in Fig.~\ref{Fig:LamK0sModel}.

The left panel of Fig.~\ref{Fig:LamK0sModel} again shows the `Soft+Quench' model predictions, while the right panel only shows two predictions from the recombination models ({\it DUKE} and {\it TEXAS}) as $\Lambda/K^{0}_{S}$ predictions only exist for these models.  
{\it OREGON} model predictions for this ratio do not exist due to the lack of knowledge of the fragmentation function for hyperons at high $p_{T}$.
Additionally, a prediction from a modified version of the {\it TEXAS} model is shown~\cite{Ref:Coalescence2}.
Note that the data have a slightly different centrality range than the range used by the models.
Also,  for the case of the `Soft+Quench' model, the predictions are for the $\Lambda/K^-$ ratio;  
this difference should have minimal effect on the comparison between data and prediction as the $K^{0}_{S}$ yield is expected to be the mean of the $K^{+}$ and $K^{-}$ yields.  
It has been shown previously that the $K^{+}/K^{-}$ ratio is dependent upon the baryon chemical potential~\cite{Ref:Barnby}, and for Au+Au collisions at $\sqrt{s_{NN}}=200~\mathrm{GeV}$, this ratio differs from unity by 10$\%$ and is independent of $p_T$~\cite{Ref:PHENIX_200GeV}.

The `Soft+Quench' results are again in good agreement with the data for $p_T~ >3~\mathrm{GeV}/c$ for the two centralities shown.  Below $3~\mathrm{GeV}/c$ the model under-predicts the data, but reproduces the general shapes of the distributions.   
As for the recombination results (right panel of Fig.~\ref{Fig:LamK0sModel}), the {\it DUKE} model, which invokes the coalescence of only thermal partons, provides reasonable agreement with the data.
The second set of model predictions are taken from the {\it TEXAS} model  and show the result for the case where soft and hard partons are allowed to coalesce and the result from a modified version of the model in which only soft, thermal partons are allowed to coalesce.
As expected, the difference between the two implementations of coalescence manifests itself for  $\sim p_T >~4~\mathrm{GeV}/c$, where the extra process leads to an enhancement above the data.
In this representation, the versions of the models which allow for the coalescence of thermal partons only agree better with the data.  
Due to the uncertainties in the fragmentation function for hyperons, the contribution to the numerator for the $\Lambda/K^{0}_{S}$  ratios from fragmentation is not included in this model, though this is only thought to be significant at higher $p_T$.  
The consequences of this is that at large $p_T$, the ratio goes to zero instead of saturating at a value of approximately 0.1.

Of the two classes of models discussed, the `Soft+Quench' model gives the best description of the data for both  light and the strange hadrons for all centralities.   
However, in order to reproduce the data, novel baryon transport techniques must be added to the model.
On the other hand, the recombination models provide an intuitive picture of particle production mechanisms for the intermediate $p_T$ range.
They also reproduce other aspects of the data, such as the scaling of the elliptic flow with valence quark number, as measured by the STAR experiment~\cite{Ref:STARv2}.
Though the different implementations describe the central data, it would be instructive to see results of calculations for more peripheral collisions taking into account the full density profile of the created medium~\cite{Ref:SystGeom}.
In spite of the agreement with data, the current implementations of the recombination mechanism can be improved.  
One of the main concerns is that entropy is not conserved due to the number of 2 $\rightarrow$ 1 and 3 $\rightarrow$ 1 processes which occur.
This particular issue could be addressed by taking resonances into account, such as the $\rho$ meson which decays into two pions.

\section{Summary and Conclusions}
We have analysed $\sqrt{s_{NN}}=200~\mathrm{GeV}$ Au+Au data taken with the STAR detector and its TPC and Ring Imaging Cherenkov Detector (RICH)
and extracted identified particle spectra as a function of centrality out to high transverse momentum ($p_T \sim 6~\mathrm{GeV}/c$).
The $\overline{p}/p$ and $\overline{\Lambda}/\Lambda$ ratios are both constant within errors as a function of $p_T$ in the measured range (up to $5~\mathrm{GeV}/c$), which is inconsistent with results from pQCD calculations up to $4~\mathrm{GeV}/c$.  At higher $p_{T}$ the interpretation is inconclusive owing to the larger error bars on the data.  
We have shown also that the $\overline{p}/\pi^{-}$ ratio rises with increasing $p_T$ up to $\sim~ 3~\mathrm{GeV}/c$, while the $\Lambda/K^{0}_{S}$ ratio increase up to  $\sim~ 3~\mathrm{GeV}/c$ and then turns over and starts to decrease.  
In both cases, the ratio exceeds unity, meaning that there is a baryon dominance over meson yields in the intermediate $p_T$ region ($\sim 2$--$5~\mathrm{GeV}/c$).
This is the first measurement to show a decrease at higher $p_{T}$ and allows us to place an upper bound on intermediate $p_{T}$ region of approximately $6~\mathrm{GeV}/c$.

We also compared the data with predictions from a variety of theoretical models.  
Even though the recombination/coalescence models represent an intuitive physical process and can reproduce well the elliptic flow~\cite{Ref:Recov2}, when more comprehensive comparisons are made to ratios of particle spectra, significant differences are seen. 
Conversely, it was found that the `Soft+Quench' model is better at reproducing both the $p_T$ and centrality dependence of the ratios, though this model contains more exotic mechanisms of baryon production (e.g. baryon junctions).  \\

\begin{acknowledgments}
We would like to thank the authors of the models discussed in this paper for providing their data points and also for many fruitful discussions.  In particular, we would like to thank R. Fries, V. Greco and I. Vitev.

We thank the RHIC Operations Group and RCF at BNL, and the NERSC Center at LBNL for their support. This work was supported in part by the HENP Divisions of the Office of Science of the U.S. DOE; the U.S. NSF; the BMBF of Germany; IN2P3, RA, RPL, and EMN of France; EPSRC of the United Kingdom; FAPESP of Brazil; the Russian Ministry of Science and Technology; the Ministry of Education and the NNSFC of China; IRP and GA of the Czech Republic, FOM of the Netherlands, DAE, DST, and CSIR of the Government of India; Swiss NSF; the Polish State Committee for Scientific  Research; STAA of Slovakia, and the Korea Sci. $\&$ Eng. Foundation.

\end{acknowledgments}

\end{document}